# Geomagnetically Induced Currents (GICs) during Strong Geomagnetic Activity (Storms, Substorms, and Magnetic Pulsations) on 23–24 April 2023

Irina Despirak[1*], Pavel Setsko[1], Andris Lubchich[1], Rajkumar Hajra[2], Yaroslav Sakharov[1], Gurbax Lakhina[3], Vasiliy Selivanov[4], Bruce Tsatnam Tsurutani[5]

[1]Polar Geophysical Institute, Apatity, Russia

[2]CAS Key Laboratory of Geospace Environment, School of Earth and Space Sciences, University of Science and Technology of China, Hefei, People's Republic of China

[3]Retired, Vashi, Navi Mumbai, India

[4]Northern Energetic Research Centre, Kola Science Center RAS, Apatity, Russia

[5]Retired, Pasadena, California, USA

[*]Corresponding author: despirak@gmail.com (I.V. Despirak)

**Abstract**

We analyzed intense geomagnetically induced currents (GICs) recorded during a complex space weather event observed on 23–24 April 2023. Two geomagnetic storms characterized by SYM-H intensities of -179 nT and -233 nT was caused by southward IMF Bz of -25 nT in the sheath fields and -33 nT in the magnetic cloud (MC) fields. GIC observations were divided into two local time sectors: nighttime (1700–2400 UT on 23 April) GICs observed during the interplanetary sheath magnetic storm, and morning sector (0200–0700 UT on 24 April) GICs observed during the MC magnetic storm. By using the direct measurements of GIC on several substations of Karelian-Kola power line (located in the north-west portion of Russia) and gas pipeline station near Mäntsälä (south of Finland), we managed to trace the meridional profile of GIC increase at different latitudes. It was shown that the night sector GIC intensification (~18-42 A) occurred in accordance with poleward expansion of the westward electrojet during substorm. On the other hand, the intense morning sector GICs (~12-46 A) were caused by Ps6 magnetic pulsations. In addition to the Ps6 pulsations, comparable in intensity morning GIC (~44 A) was associated with a local substorm-like disturbance caused by a high density solar wind structure, possibly a coronal loop portion of an ICME.

## 1. Introduction

It is known that during geomagnetic disturbances electric fields are induced, which in turn can create intense, low-frequency (~0.001–1 Hz), quasi-direct currents in terrestrial technological networks, called geomagnetically induced currents (GICs; Barlow et al., 1849; Varley, 1873; Campbell, 1980; Akasofu & Aspnews, 1982; Viljanen et al., 2006; Pulkkinen et. al. 2005; Lakhina et al., 2021; Tsurutani and Hajra, 2023). It has been established that the most severe impact of geomagnetic disturbances on technological systems are GICs, associated with sudden changes in the geomagnetic field (dB/dt > 1 nT s$^{-1}$; Pilipenko, 2023). Rapid changes in the geomagnetic field may be associated with interplanetary shocks or solar wind plasma parcel impingements upon the Earth's magnetosphere (see Tsurutani & Hajra, 2021, and references therein). GICs are the final link in the communication chain: Sun – solar wind – magnetosphere – ionosphere – Earth's surface (Tsurutani et al., 2021). At the same time, the intensity of GICs depends not only on the intensity of magnetic disturbances, but also on the location and configuration of the technological system where such currents are induced (e.g. Lui et al., 2018; Clilverd et al., 2021). In present day technology, lengthy power lines are increasingly being built to transmit large volumes of energy over long distances, and GIC levels have become 2–3 times higher than those observed 20 years ago during magnetic storms of the same intensity (Molinski, 2002).

Control over the excitation of GICs in technological systems is carried out in many countries located in both high and middle geomagnetic latitudes (e.g. Kelly et al., 2017; Mac Manus et al., 2017; Clilverd et al., 2021; Watari et al., 2021). In the north-west of Russia, the Polar Geophysical Institute together with Northern Energetic Research Centre in the frame of EC grant EURISGIC (2011-2014) built a system for monitoring such currents in the solidly grounded neutral wire of autotransformers in the existing Karelian-Kola power transmission line (Viljanen, 2011; Viljanen et al., 2012; Sakharov et al., 2007; 2016; Selivanov et al., 2023).

The GIC variations at substations in north-west Russia has been studied since 2011, and it was found by Sakharov et al. (2019) that intense GICs were associated mainly with interplanetary coronal mass ejections (ICMEs) and related phenomena (interplanetary sheaths). Similar results were obtained for GICs recorded on the Mäntsälä gas pipeline. It was shown that 91% of days with GIC peaks > 10 A were associated with ICMEs (Huttunen et al., 2008). Based on an analysis of the Mäntsälä GIC data from 1999 to 2019, it was found that ~51% of GIC > 10 A peaks are associated with interplanetary sheaths, ~44% with magnetic clouds (MCs), and ~3% with corotating interaction regions (CIRs; Tsurutani et al., 2021; Hajra, 2022a). Intense GICs were observed during intense and long-duration auroral substorms (Hajra, 2022b, Despirak et al., 2022).

In addition, the GIC increase can be influenced by local disturbances of the geomagnetic field associated with the development of magnetic pulsations, omega band structures in auroras, etc. (Oliveira et al., 2017; Vorobjev et al., 2018; Apatenkov et al., 2019; Kozyreva et al., 2020; Lakhina et al., 2020; Yagova et al., 2021; Sakharov et al., 2022). Recently it has been established that the main sources of GIC growth at auroral zone latitudes are the intensification and poleward movement of the westward electrojet current during substorm expansion phases (Despirak et al., 2022 and 2023) as well as Pc5 pulsations, usually observed during the recovery phase of substorms (Setsko et al., 2023).

Note that in the absence of GIC registration data, their intensities are studied using ground-based magnetometer data by identifying rapid geomagnetic field changes (dB/dt), so-called "spikes" (e.g., Ngwira et al., 2018; Dimmok et al., 2019; Schillings et al., 2022). Large dB/dt bursts have been shown to be controlled by solar activity and have been observed predominantly in pre-midnight, morning and pre-noon magnetic local times (MLTs) (Milan et al., 2023). Milan et al. (2023) suggested that it is possible that these large bursts are associated with the substorm onsets (pre-midnight sector), omega bands (morning sector), and Kelvin-Helmholtz instability (pre-noon sector). Moreover, other studies of GIC increase during very intense substorms, so called supersubstorms (SSS) (Tsurutani et al., 2015) showed that large dB/dt bursts occurred during the SSS periods in the latitude band of 60°–75°.

The aim of the present work is to study of sources GICs during the space weather event occurring on 23–24 April 2023. Extremely intense GICs (up to 46 A) were recorded at substations in the north-west of Russia and in Finland.

## 2. Data

At this work we use following data. In the frame of EC grant EURISGIC (2011-2014), the Polar Geophysical Institute, together with the Northern Energetic Research Centre of the KSC RAS, created and subsequently maintain the system for monitor GICs in the existing Karelian-Kola power transmission line with a length of over 800 km (http://eurisgic.ru/). The network includes five stations: "Vykhodnoy", "Revda", "Titan" (Murmansk region), "Loukhi" and "Kondopoga" (Republic of Karelia). At each substation uses Hall sensors that directly measure the currents flowing into the ground through the grounded neutral wire of autotransformers. Positive values mean GICs going into the ground (Sakharov et al., 2007; 2016; Barannik et al., 2012; Selivanov et al., 2023). We use also the data from the east-west aligned gas pipeline station located near Mäntsälä, Finland (https://space.fmi.fi/gic/index.php).

Figure 1 shows the scheme of the GIC recording points (red and white dots with data and no data at selected time, respectively) and the location of the IMAGE magnetometers (black dots). Geographical coordinates of the substations, the data of which were used in the work: Vykhodnoy (VKH) (68.8° N, 33.1° E), Kondopoga (KND) (62.2° N, 34, 3°E), Mäntsälä (MAN) (60.6° N, 25.2° E). The whole list of the coordinates and names used in this work IMAGE magnetometers and GIC registration stations are given in Table 1. It can be seen that the observation zone occupies a wide area in latitude, extending from subauroral to high latitudes, and GIC registration occurs at geographic (geomagnetic) latitudes from ~60° to ~69° (from ~57.3° to 65.5°). Note that this location corresponds to latitudes where substorm disturbances are usually observed.

Figure 1

The spatial distribution of the substorm electrojets was determined using the magnetometers of the IMAGE (http://space.fmi.fi/image/) and SuperMAG (http://supermag.jhuapl.edu/) networks (Gjerloev, 2009; Newell and Gjerloev, 2011). The onsets of substorms have been determined by sharp onsets of negative bays in the magnetic X-component of the most equatorial station of the meridional chain PPN-NAL. Maps of the distribution of ionospheric equivalent currents were taken from the MIRACLE website (https://space.fmi.fi/MIRACLE) (Viljanen and Häkkinen, 1997).

The IL-index was taken from IMAGE network and shows the variation of the magnetic field at the selected IMAGE stations, that is, it is similar to the AL index. The PC-index was taken from the International Service of Geomagnetic indices (ISGI, https://isgi.unistra.fr).

The solar wind and IMF parameters, SYM/H and ASY/H indexes were taken from the CDAWeb data service (https://cdaweb.gsfc.nasa.gov/).

## 3. Solar wind conditions and geomagnetic activity

Solar wind and interplanetary magnetic field (IMF) parameters for the period from 00 UT on April 23 to 24 UT on April 24, 2023 are given in Figure 2. The figure shows from top to bottom: the magnitude ($B_T$) of the IMF; the Y- and Z-component of the IMF ($B_Y$, $B_Z$); the flow speed (V), the density (N), temperature (T), and dynamic pressure (P) of solar wind, and the geomagnetic indices PC, ASYM/H and SYM/H. On April 21, 2023, at ~20:40 – 21:40, an M1.7 class solar flare occurred. The flare location was almost exactly on the Sun-Earth line. After a powerful flare, a huge prominence broke off from the Sun, forming a portion of a coronal mass ejection (CME: Illing and Hundhausen, 1986). It should be noted that the prominence did not reach the Earth with the rest of the interplanetary CME (ICME). The ICME reached the Earth on April 23 at ~17:30 UT and an interplanetary shock (IS) was detected, followed by the sheath region and magnetic cloud (MC). The shock had a Mach number of 2.7, and was quasiperpendicular: $\theta_{kB}$ = 81 deg. The sheath was present from ∼17:30 UT on April 23 to ∼01:50 UT on April 24, and the MC period lasted from ~01:50 UT to ~22:20 UT on 24 April. The boundaries are marked by the red rectangles and inscription "Sheath" and "MC", respectively. It can be seen that there were two following periods $B_Z$<0 during the sheath ($B_Z$~-25 nT) and at the beginning of the MC ($B_Z$~-33 nT). The first southward $B_Z$ is caused by shock compression of preexisting southward Bz (Tsurutani et al. 1988). These two periods $B_Z$ < 0 led to the development of two strong magnetic storms. The first storm (with peak SYM/H of – -179 nT) was observed during the Sheath; the second storm occurred during MC (with peak SYM/H of – -233 nT). On April 23, 2023, from ~18:00 UT to 21:00 UT, there were three substorms registered by IMAGE chain, blue vertical lines indicate their onsets. But note, that Figure 2 presents the moments of substorm onsets rather schematically since the time interval is quite long.

Figure 2

More details on the magnetic storm development and geomagnetic activity are presented in Figure 3. The top panels shown the variations of SYM/H and IL indexes, the bottom panels the variations of GICs at substations Vykhodnoy (VKH) and Kondopoga (KND). It is seen that before midnight there were 4 substorms (IL ~ –515 nT, ~ –1512 nT, ~ –1345 nT and ~ –574 nT). This sector is marked by the green rectangle in Figure 3. This period is associated with the first magnetic storm (~17:30 – 24 UT on 23 April). During second storm, the intense geomagnetic activity was observed during ~02 – ~07 UT on 24 April connected to geomagnetic pulsations and one intense substorm-like disturbance (peak of IL ~ –1900 nT) at ~04:00 UT on 24 April. These two time sectors are marked by green rectangles and are designated the "First interval" and "Second interval", respectively. During this complicated magnetic activity strong GICs were registered at MAN, KND and VKH. The largest GICs 42 A were observed at ~ 20:30 UT and -42 A at ~20:45 UT on 23 April, 46 A at 03:35 UT and 44 A at ~04:00 UT on 24 April, 2023. The entire data analysis was divided into these two intervals. The First interval was associated with the study of GICs observed in the premidnight sector, during the sheath storm. The Second interval was observed in the morning sector, during the MC storm. The GICs that occurred during these two intervals on the Karelian-Kola power line (VKH and KND stations) and on the Mäntsälä (MAN station) pipeline will be considered in more detail below.

Figure 3

## 4. Observations

### a) First interval

The magnetic disturbances and GICs registrations for the First interval, from 16 to 24 UT on April 23 are shown on Figure 4. The left panel shows the variations of X component of IMAGE magnetometers from Tartu (TAR) to Ny Ålesund (NAL), TAR-NAL chain (a), the maps of latitudinal profile of the westward and eastward electrojets, calculated by the MIRACLE system, are located on the upper right panel (b). The location of GICs substations (VKH, KND, MAN) are marked by horizontal red dashed lines. Below this panel, there are GIC profiles (red lines) with the corresponding magnetic components of the nearest IMAGE station (black lines) (c). For these nearest stations – Lovozero (LOZ) is close to VKH, Hankasalmi (HAN) is close to KND and Nurmijärvi (NUR) is close to MAN –variations of X- and Y component of magnetic field are presented. On this two panels the geographical latitudes (GLAT) are used. Three periods of different magnetic disturbances and large values of GIC (17:30 – 18:30 UT, 19:00 – 20:10 UT, 20:20 – 21:00 UT) are marked by horizontal blue, green and purple bars, respectively.

Figure 4

According to variations in the X-component, the disturbances began at ~17:40 UT (corresponding to the moment of arrival of the IS shock wave at the Ranua (RAN) - Ny Ålesund (NAL) chain stations simultaneously, with the westward electrojet observed only at high-latitude stations: Bear Island (BJN) - Ny Ålesund (NAL), and the eastward electrojet observed at lower latitudes. Registration of negative bays only at high latitudes indicates the development of a so-called polar substorm at ~17:40 UT (Despirak et al., 2018; Kleimenova et al., 2023). The onset of the substorm was near the SOR station, and strong variations in the Y and Z components were recorded there (not shown in Figure 4 for brevity). As a result, very strong GICs (~-32 A) at ~ 17.45 UT were recorded at the VKH station, close to the onset of the polar substorm. While at the lower latitude stations KND and MAN only small GICs (~2-3 A) were detected, which corresponded to magnetic disturbances associated with the arrival of the magnetospheric magnetosonic wave induced by the interplanetary shock wave.

Rapid movements of the westward and eastward electrojets from high to lower latitudes (from 75° to 62° GLAT) was observed from ~18:30 to ~19 UT, during the growth phase of the second substorm. At ~19:00 UT the second substorm began, with sharp negative bays recorded at the TAR to RAN stations. As a result, intense GICs at ~19:09 UT occurred at the KND (~8A) and MAN (~27A) stations, close to the substorm onset. At the VKH station, intense GICs (~30A) were also recorded, which may have been caused by a sharp movement of the electrojet towards the equator during the substorm growth phase.

The next intense substorm began at ~19:30 UT, disturbances were observed in a wide region from TAR (~58°GLAT) to NAL (~75°GLAT). Figure 4b shows the development of equivalent westward (blue color) and eastward (yellow color) currents. It is seen the westward electrojet jumped poleward during the expansion phase of this substorm. The electrojet started at ~19:30 UT at NUR, then reached SOR at ~19:45 UT and finally appeared at BJN at ~20:08 UT. The expansion phase of the substorm is marked by a horizontal green line in Figure 4c. It is also noted that the substorm consisted of 3 intensifications recorded at SOR at ~19:45, ~19:52 and ~20:05 UT (Figure 4a). Accordingly the spatial-temporal structure of the westward electrojet development were related to GIC peaks observed at MAN (~27 A) and KND (~14 A) at ~19:30 UT, and at VKH (~18 A, ~18 A, ~32 A) at ~19:46, ~19:56 and ~20:09 UT.

The strongest GIC (-42 A) was observed at VKH at ~20:43 UT, when there were large negative (~ –1500 nT) and positive (~750 nT) bays in the X- and Y- components in LOZ, respectively (Figure 4c). Note that, due to location of the Karelian-Kola power line and by Faraday's law of induction, increase of the GIC at VKH and KND stations will be mainly driven by rapid change in the Y-component of the magnetic field (Setsko et al., 2023). As seen from Figures 4a and 4b, these strong magnetic disturbances were associated with the appearance of a new substorm during the recovery phase of the previous substorm. The new substorm interval is indicated by a purple horizontal line at the Figure 4c. Negative bays began at ~20:30 UT on NUR station. The bays appeared at SOR at ~ 20:45 UT, and were even stronger at PEL and RAN (Figure 4a). The most intense westward electrojet was observed ~67° GLAT (Figure 4b). Corresponding GICs were observed at ~20:30 UT at MAN (~12A) and at KND (~8A).

Figure 5 shows the solar wind and IMF conditions before onset of third substorm observed at the recovery phase of second substorm, i.e., from 20:00 to 20:40 UT on April 23. Variations of the magnitude $B_T$, $B_X$, $B_Y$ and $B_Z$ components of the IMF and proton density ($N_P$) are shown. Note that Figure 5 is obtained by data from the Wind satellite. The propagation time delay from WIND to Earth was ~35 min. It is seen that $B_Z$ component turned northward at ~20.12 UT that possibly

[Figure 5]

triggered the third substorm, which included large variations of the Y-component of the geomagnetic field. It resulted in the highest recorded GIC (~42A) at Vykhodnoy station for first interval, from 17 to 24 UT on 23 April, 2023.

**b) Second interval**

The magnetic disturbances and GICs for the Second interval, from 00 to 08 UT on April 24, 2023 are shown on Figure 6. The format of Figure 6 is similar to that of Figure 4. Strong disturbances of the magnetic field and the intense GIC were registered from ~ 02 UT to ~07 UT on April 24. Note that this period corresponded to the second storm which was caused by the prolonged period of southward $B_Z$ of the IMF (~ –33 nT) during the MC (Figure 2). The minimum SYM/H (~ –233 nT) occurred at ~03 – 05 UT on 24 April. The interval of intense GICs corresponded to main phase of the second magnetic storm. As seen from Figure 6a and 6b, the geomagnetic pulsations were observed during time period from ~02 UT to ~07:30 UT. From the Finnish induction magnetometer data, Pi1 pulsations were registered at auroral stations, moreover the amplitude of these pulsations was larger at NUR station (60.5°GLAT) then at SOD station (67.3°GLAT) (not shown for brevity). During the same period, from ~02:30 to ~07:00 UT, in addition to irregular Pi1 pulsations, Ps6 pulsations with periods ~20–30 min are also noted in the magnetograms (Figure 6). As a result, intense GIC peaks occurred at many ground stations: from ~ 02 to ~06 UT at MAN (~10–35 A), from ~ 03 to ~05:40 UT at KND (~7–15 A), and from ~ 03 to ~07:20 UT at VKH (~15–45 A).

Note that the strongest GICs at VKH (~45 A) were during periods when strong disturbances were observed in both the LOZ station X- and Y- components, at ~03:30-03:50 UT and at~ 04:00 UT. Stronger GICs at KND were -9 A at ~03:15 and -15 A at ~03:53 UT, correspondingly to peaks in Y-component at HAN. A strong GIC at MAN (-35 A) occurred at~ 03:56 UT which corresponded to the strongest peak in the X- component at the NUR. The period of the strong magnetic disturbances and the corresponding largest GICs (03:50 – 04:10 UT) is marked by a horizontal blue line (Figure 6c).

Figure 6

It is interesting to note that the strongest GIC occurred almost simultaneously at ~04:00 UT at all stations (VKH, KND, MAN) and were associated with maximum negative and positive bays in the X- and Y-components. We suggest that the GICs are not due to magnetic pulsations, but to other phenomena that occurred during this period of time interval. Figure 7a presented the solar wind and IMF conditions for short period from 02:50 UT to 03:20 UT on April 24, 2023. The format of Figure 7a is similar to that of Figure 5. Note that Figure 7a is obtained by WIND data with a propagation delay of ~37 min. It is seen that at ~03:02 UT the density increased from ~5 to 45 protons/cm$^3$. The density jump was accompanied by the decrease in the magnitude $B_T$ from 33.15 to 30.0 nT. This structure is a coronal loop propagated to 1 au (Galvin et al., 1987; Tsurutani et al. 1998). This is the first time that a coronal loop portion of an ICME has been found to be geoeffective.

Figure 7

Figure 7b shows variations of components of the magnetic field at some SuperMAG stations, located at Siberia (IRT, NVS), the Urals (ARS) and the north of the Karelia region (WSE). The local strong negative bay was observed at all stations: at ~03:45 UT at IRT, NVS, ARS, at ~04:00 UT at WSE. The disturbance was very local, and by duration, a prolonged ~15 minutes. Note that at WSE, located in the north of the Karelia region, this local disturbance was observed during strong geomagnetic pulsations (similar to the IMAGE stations). It can be seen that the strong GICs during Second interval on April 24 were also observed at ~03:50–04:00 UT, when the local substorm-like disturbance caused by the structure of the high-density solar wind/coronal loop was additionally superimposed.

**4. Discussion**

We studied intense GICs (~12-46A) detected during an unusual ICME event. Although the interplanetary event consisted of a fast magnetic cloud preceded by a shock and sheath, there was a high density region between the sheath and the magnetic cloud which appears to be a coronal loop. The loop is the first one detected at 1 au that has been found to be geoeffective. The sheath negative Bz fields caused a magnetic storm of -179 nT and the magnetic cloud negative Bz fields caused a second magnetic storm of -233 nT intensity.

The main features of these two storms of interest in this paper are that the strong GICs observed in the pre-midnight for one storm and post-midnight (morning) sector for the other storm. The first storm was caused by a long period of southward $B_Z$ in the sheath preceding the MC. The second magnetic storm was caused by a long interval of the southward $B_Z$ within a MC. At the same time, the region of north-west Russia and southern Finland, where the GICs registration stations are located, was in the night and morning sectors during the first and second storms, respectively.

The first case of GIC observations during the first storm (from 17-24 UT) refers to the MLT night sector (20–03 MLT). For the second storm (02–07 UT) the stations were in the morning and noon sectors (05–11 MLT). As can be seen from Figures 4 and 6, large bursts of GICs (~10–45 A) occurred during both intervals. In the first interval, the sources of the GICs were magnetic disturbances related to a substorm, during which an intensification of the westward electrojet and its poleward expansion were observed.

For the second case, strong GICs were associated with Ps6 geomagnetic pulsations. As was shown recently, strong bursts (dB/dt) of magnetic field disturbances were distributed over different MLT sectors (in the pre-midnight, morning and daytime sectors). These GICs were associated with the onsets of a substorm (pre-midnight sector), omega bands (morning sector), and Kelvin-Helmholtz instability (pre-noon sector) (Milan et al. 2023). It is known that Ps6 magnetic pulsations are observed in the morning sector, during the substorm recovery phase, and are considered a magnetic signature of spatially periodic optical auroras, known as omega bands (Safargaleev et al., 2005). The bright band auroral patches are caused by electron precipitation, and the set of Ps6 pulsations are generated by the associated current systems (Jorgensen et al., 1999). That is, the source of strong magnetic disturbances in the morning sector are omega bands or Ps6 pulsations, which can lead to strong GICs. Our observations confirmed that source of intense GICs during the second interval, in the morning sector, were Ps6 pulsations. It was also shown that the source of strong GICs observed in the night sector during the first interval was associated with the onset of the substorm and its further development (the poleward expansion during the expansion phase). Our results are in agreement with Milan et al. (2023) that GICs sources can be divided into MLT sectors: in the pre-midnight sector these will be phenomenon connected with substorms, in the morning sector, with geomagnetic pulsations.

The use of the extended meridional profile of the GICs registration stations made it possible to trace the increase of the intense GIC in accordance with the poleward expansions of substorms. GIC intensifications at different stations (MAN, KND, VKH) occurred simultaneously with the movement of the westward electrojet during the substorm expansion phase.

It should be noted that in accordance with the geographic location of the GICs registration chain and Faraday's law of induction, increase of the GICs at VKH and KND stations will be mainly driven by rapid change in the Y-component of the magnetic field, while at the MAN station in the X-component (Setsko et al., 2023). It was shown that the strongest GIC (~42A) during first interval, observed at ~20:45 UT on 23 April.2023, was related to the appearance of a new substorm intensification caused by the rotation of the IMF $B_Z$ to the north (Figures 4 and 5). Unlike the previous ones, this substorm was characterized by a strong positive disturbance in the Y-component, which led to the development of the strongest GIC during the first storm.

During second interval, the strongest GICs (~45 A) was registered in all stations simultaneously at ~03:50 – 04:00 UT on 24 April (Figure 6). It was shown that it was connected with the local substorm-like intensification with intense pulsations. This local disturbance has been registered at magnetometers located in Siberia (IRT, NVS), the Urals (ARS), in the north of north of the Karelia region (WSE) and in the IMAGE magnetometers chain (Figure 6 and 7b). We believe that such local substorm-like intensifications may be caused by solar wind pressure pulses such as interplanetary shocks (Tsurutani and Hajra, 2023) and plasma parcels (Tsurutani and Hajra, 2015).

We would also like to note that our paper is in agreement with Akasofu and Aspnews (1982) that GICs could be related to substorm activity, but provides more detailed information.

5. Conclusions

The sources of intense GICs (~12–46A) detected during two magnetic storms on 23–24 April, 2023 in electrical circuits in the north-west of Russia and on gas pipeline in Finland have been identified. Intense bursts of GIC occurred both at the first and second magnetic storms. For the first storm, in the night MLT sector, the source of the GIC was substorm development, while at the second storm, in the morning MLT sector, it was Ps6 geomagnetic pulsations.

- GIC events occurred simultaneously with poleward expansions of the westward electrojet during the expansion phase of substorms. The strongest GIC (~42A) during first storm, observed at ~20:45 UT on 23 April.2023, has been related to the appearance of a new substorm intensification characterized by a strong positive disturbance in the Y- component.

- It was established also that intense GICs (~15–46 A) in the morning sector were caused by Ps6 magnetic pulsations. But the comparable in intensity GIC (~45 A) was associated with a local substorm-like intensification caused by the magnetospheric impact of a high density structure in the solar wind (see also Tsurutani and Hajra for such cases causing supersubstorms). This plasma structure was the coronal loop portion of an ICME. This is the first time that a coronal loop has been shown to be geoeffective.

**Data Availability**

Datasets related to this article can be found at:

http://cdaweb.gsfc.nasa.gov/cdaweb/istp_public/ (Solar wind and IMF data); https://space.fmi.fi/image/ (IMAGE magnetometers network, Mäntsälä GIC data as well as IL-index);

https://isgi.unistra.fr (PC-index);

https://eurisgic.ru/ (GICs data on Karelian-Kola power line)

**Acknowledgements**

The authors are grateful to the creators of the OMNI databases (http://cdaweb.gsfc.nasa.gov/cdaweb/istp_public/), the SuperMAG network (http://supermag.jhuapl.edu/), the MIRACLE system (https://space.fmi.fi/MIRACLE/) for the ability to use them in our work. And also to the Finnish Meteorological Institute for Mäntsälä GIC data, IL-index and IMAGE magnetometers network (https://space.fmi.fi/image/).

# References

Akasofu, S-I, Aspnes, JD. 1982. Auroral effects on power transmission line systems. *Nature*, **295**: 136–137. https://doi.org/10.1038/295136a0.

Apatenkov, SV, Pilipenko, VA, Gordeev, EI, Viljanen, A, Juusola, L, Belakhovsky, VB, Sakharov, YaA, Selivanov, VN. 2020. Auroral omega bands are a significant cause of large geomagnetically induced currents. *Geophys. Res. Lett.*, **47**: e2019GL086677. https://doi.org/10.1029/2019GL086677.

Barannik, MB, Danilin, AN, Kat'kalov, YuV, Kolobov, VV, Sakharov, YaA, Selivanov, VN. 2012. A system for recording geomagnetically induced currents in neutrals of power autotransformers. *Instruments and Experimental Techniques*, **55**: 110–115. https://doi.org/10.1134/S0020441211060121.

Barlow, WH, Barlow, P, Culley, RS. 1849. VI. On the spontaneous electrical currents observed in the wires of the electric telegraph. *Philosophical Transactions of the Royal Society of London*, **139**: 61–72. https://doi.org/10.1098/rstl.1849.0006.

Campbell, WH. 1980. Observation of electric currents in the Alaska oil pipeline resulting from auroral electrojet current sources. *Geophysical Journal International*, **61**: 437–449. https://doi.org/10.1111/j.1365-246X.1980.tb04325.x.

Clilverd, MA., Rodger, CJ, Freeman, MP, Brundell, JB, Mac Manus, DH, et al. 2021. Geomagnetically induced currents during the 07–08 September 2017 disturbed period: a global perspective. *J. Space Weather Space Clim.*, **11**: Art. 33. https://doi.org/10.1051/swsc/2021014.

Despirak, IV, Lubchich, AA, Kleimenova, NG. 2018. High-latitude substorm dependence on space weather conditions in solar cycle 23 and 24 (SC23 and SC24). *J. Atm. Solar-Terr. Phys.*, **177**: 54–62. https://doi.org/10.1016/j.jastp.2017.09.011.

Despirak, IV, Setsko, PV, Sakharov, YaA, Lyubchich, AA, Selivanov, VN, Valev, D. 2022. Observations of geomagnetic induced currents in Northwestern Russia: case

studies. *Geomagnetism and Aeronomy*, **62**: 711–723. https://doi.org/10.1134/S0016793222060032.

Despirak, IV, Setsko, PV, Sakharov, YaA, Lubchich, AA, Selivanov, VN. 2023. Geomagnetically induced currents during supersubstorms on September 7-8, 2017. *Bulletin of the Russian Academy of Sciences: Physics*, **87**: 999–1006. https://doi.org/10.3103/S1062873823702283.

Dimmock, AP, Rosenqvist, L, Hall, J-O, Viljanen, A, Yordanova, E, Honkonen, I, André, M, Sjöberg, EC. 2019. The GIC and geomagnetic response over Fennoscandia to the 7–8 September 2017 geomagnetic storm. *Space Weather*, **17**: 989–1010. https://doi.org/10.1029/2018SW002132.

Galvin, AB, Ipavich, FM, Gloeckler, G, Hovestadt, D, Bame, SJ, Klecker, B, Scholer, M, Tsurutani, BT. 1987. Solar wind iron charge states preceding a driver plasma. *J. Geophys. Res.*, **92**: 12069–12081. https://doi.org/10.1029/JA092iA11p12069.

Gjerloev, JW. 2009. A Global Ground-Based Magnetometer Initiative. *Eos Trans. AGU*, **90**: 230–231. https://doi.org/10.1029/2009EO270002.

Hajra, R. 2022. Intense geomagnetically induced currents (GICs): Association with solar and geomagnetic activities. *Sol. Phys.*, **297**: 14. https://doi.org/10.1007/s11207-021-01945-8.

Huttunen, KEJ, Kilpua, SP, Pulkkinen, A, Viljanen, A, Tanskanen, E. 2008. Solar wind drivers of large geomagnetically induced currents during the solar cycle 23. *Space Weather*, **6**: S10002. https://doi.org/10.1029/2007SW000374.

Kelly, GS, Viljanen, A, Beggan, CD, Thomson, AWP 2017. Understanding GIC in the UK and French high-voltage transmission systems during severe magnetic storms. *Space Weather*, **15**: 99–114. https://doi.org/10.1002/2016SW001469.

Jorgensen, AM, Spence, HE, Huges, TJ, McDiarmid, D. 1999. A study of omega bands and Ps6 pulsations on the ground, at low altitude and at geostationary orbit. *J. Geophys. Res.*, **104**: 14 705–14 715. https://doi.org/10.1029/1998JA900100.

Kleimenova, NG, Despirak, IV, Malysheva, LM, Gromova, LI, Lubchich, AA, Roldugin, AV, Gromov, SV. 2023. Substorms on a contracted auroral oval. *J. Atm. Solar-Terr. Phys.*, **245**: 106049. https://doi.org/10.1016/j.jastp.2023.106049.

Kozyreva, O, Pilipenko, V, Krasnoperov, R, Baddeley, L, Sakharov, Y, Dobrovolsky, M. 2020. Fine structure of substorm and geomagnetically induced currents. *Ann. Geophys.*, **63**: GM219. https://doi.org/10.4401/ag-8198.

Lakhina, GS, Hajra, R, Tsurutani, BT. 2021. Geomagnetically induced currents. In H. K. Gupta (Ed.), Encyclopedia of solid earth geophysics (pp. 523–527). Cham: Pringer International Publishing. https://doi.org/10.1007/978-3-030-58631-7_245.

Liu, CM, Wang, X, Wang, H, Zhao, H. 2018. Quantitative influence of coast effect on geomagnetically induced currents in power grids: a case study. *J. Space Weather Space Clim.*, **8**: A60. https://doi.org/10.1051/swsc/2018046.

Mac Manus, DH, Rodger, CJ, Dalzell, M, Thomson, AWP, Clilverd, MA, Petersen, T, Wolf, MM, Thomson, NR, Divett, T. 2017. Long-term geomagnetically induced current observations in New Zealand: Earth return corrections and geomagnetic field driver. *Space Weather*, **15**: 1020–1038. https://doi.org/10.1002/2017SW001635.

Milan, SE, Imber, SM, Fleetham, AL, Gjerloev, J. 2023. Solar cycle and solar wind dependence of the occurrence of large dB/dt events at high latitudes. *J. Geophys. Res. Space Phys.*, **128**: e2022JA030953. https://doi.org/10.1029/2022JA030953.

Molinski, TS. 2002. Why utilities respect geomagnetically induced currents. *J. Atmos. Terr. Phys.*, **64**: 1765–1778. https://doi.org/10.1016/S1364-6826(02)00126-8.

Ngwira, CM, Sibeck, D, Silveira, MV, Georgiou, M, Weygand, JM, Nishimura, Y, Hampton, D. 2018. A study of intense local dB/dt variations during two geomagnetic storms. *Space Weather*, **16**: 676–693. https://doi.org/10.1029/2018SW001911.

Newell, PT, Gjerloev, JW. 2011. Evaluation of SuperMAG auroral electrojet indices as indicators of substorms and auroral power, *J. Geophys. Res.*, 116: A12211. https://doi.org/10.1029/2011JA016779.

Oliveira, DM, Ngwira, CM. 2017. Geomagnetically induced currents: Principles, *Braz. J. Phys.*, **47**: 552–560. https://doi.org/10.1007/s13538-017-0523-y.

Pilipenko, VA, Chernikov, AA, Solovjev, AA, Yagova, NV, Sakharov, YA, Kudin, DV, Kostarev, DV, Kozyreva, OV, Vorobev, AV, Belov, AV. 2023. Influence of space weather on the reliability of the transport system functioning at high latitudes. *Rus. J. Earth Sci.*, **23**: ES2008. https://doi.org/10.2205/2023es000824 (in Russian).

Pulkkinen, A, Lindahl, S, Viljanen, A, Pirjola, R. 2005. Geomagnetic storm of 29–31 October 2003: Geomagnetically induced currents and their relation to problems in the Swedish high-voltage power transmission system. *Space Weather*, **3**: S08C03. https://doi.org/10.1029/2004SW000123.

Safargaleev, V, Sergienko, T, Nilsson, H, Kozlovsky, A, Massetti, S, Osipenko, S, Kotikov, A. 2005. Combined optical, EISCAT and magnetic observations of the omega bands/Ps6 pulsations and an auroral torch in the late morning hours: a case study. Annales Geophysicae, **23**: 1821–1838. https://doi.org/10.5194/angeo-23-1821-2005.

Sakharov, YaA, Danilin, AN, Ostafiichuk, RM. 2007. Recording of GICs in power systems of the Kola Peninsula, in Trudy 7-go Mezhdunar. simp. po elektromagnitnoi sovmestimosti i elektromagnitnoi ekologii (Proceedings of the 7th International Symposium on Electromagnetic Compatibility and Electromagnetic Ecology), St. Petersburg: IEEE, pp. 291–293.

Sakharov, YaA, Katkalov, YuV, Selivanov, VN, Viljanen, A. 2016. Recording of GICs in a regional power system, in Prakticheskie aspekty geliogeofiziki, Materialy spetsial'noi sektsii "Prakticheskie aspekty nauki kosmicheskoi pogody" 11-i ezhegodnoi konferentsii "Fizika plazmy v solnechnoi sisteme" (Practical Aspects of Heliogeophysics: Proceedings of the Special Section "Practical Aspects of the Science of Space Weather" of the 11th Annual Conference "Physics of Plasma in the Solar System"), Moscow. IKI, pp. 134–145.

Sakharov, YaA, Selivanov, VN, Bilin, VA, Nikolaev, VG. 2019. Extremal values of geomagnetically induced currents in the regional power system. "Physics of Auroral Phenomena", Proc. XLII Annual Seminar, Apatity, pp. 53-56. DOI: 10.25702/KSC.2588-0039.2019.42.53-56 (in Russian).

Sakharov, YaA, Yagova, NV, Pilipenko, VA, Selivanov, VN. 2022. Spectral content of Pc5–6/Pi3 geomagnetic pulsations and their efficiency in generation of geomagnetically induced currents. *Russ. J. Earth Sci.*, **22**: ES1002. https://doi.org/10.2205/2021ES000785.

Selivanov, VN, Aksenovich, TV, Bilin, VA, Kolobov, VV, Sakharov, YA. 2023. Database of geomagnetically induced currents in the main transmission line "Northern transit". *Solar- Terr. Phys.*, **9**: 93101. https://doi.org/10.12737/stp-93202311.

Schillings, A, Palin, L, Opgenoorth, HJ, Hamrin, M, Rosenqvist, L, Gjerloev, JW, Juusola, L, Barnes, R. 2022. Distribution and occurrence frequency of dB/dt spikes during magnetic storms 1980–2020. Space Weather, **20**: e2021SW002953. https://doi.org/10.1029/2021SW002953.

Setsko, PV, Despirak, IV, Sakharov, YaA, Lubchich, AA, Bilin, VA, Selivanov, VN. 2023. Geoinduced currents on Karelian-Kola power line and Finnish gas pipeline on September, 12–13 2017. *J. Atm. Solar-Terr. Phys.*, **247:** 106079. https://doi.org/10.1016/j.jastp.2023.106079.

Tsurutani, BT, Gonzalez, WD, Tang, F, Akasofu, S-I, Smith, EJ. 1988. Origin of interplanetary southward magnetic fields responsible for major magnetic storms near solar maximum (1978-1979). *J. Geophys. Res.*, **93**: 8519-8531. https://doi.org/10.1029/JA093iA08p08519.

Tsurutani, BT., Arballo, JK, Lakhina, GS, Ho, CM, Ajello, J, et al. 1998. The January 10, 1997 auroral hot spot, horseshoe aurora and first substorm: A CME loop? *Geophys. Res. Lett.*, **25**: 3047–3050. https://doi.org/10.1029/98GL01304.

Tsurutani, BT, Lakhina, GS, Verkhoglyadova, OP, Gonzalez, WD, Echer, E, Guarnieri, FL. 2011. A Review of interplanetary discontinuities and their geomagnetic effects. *J. Atm. Solar-Terr. Phys.*, **73**: 5–19. https://doi.org/10.1016/j.jastp.2010.04.001.

Tsurutani, BT, Hajra, R, Echer, E, Gjerloev, JW. 2015. Extremely intense (SML ≤ –2500 nT) substorms: Isolated events that are externally triggered? *Ann. Geophys.*, **33**: 519–524. https://doi.org/10.5194/angeo-33-519-2015.

Tsurutani, BT, Hajra, R. 2021. The Interplanetary and Magnetospheric causes of Geomagnetically Induced Currents (GICs) > 10 A in the Mäntsälä Finland Pipeline: 1999 through 2019. *J. Space Weather Space Clim.*, **11**: 23. https://doi.org/10.1051/swsc/2021001.

Tsurutani, BT, Hajra, R. 2023. Energetics of Shock-triggered Supersubstorms (SML < –2500 nT). *ApJ*, **946**: 17. https://doi.org/10.3847/1538-4357/acb143.

Varley, C. 1873. Discussion of a few papers on earth currents. *Journal of the Society of Telegraph Engineers*, **2**: 111–123. https://doi.org/10.1049/jste-1.1873.0033.

Viljanen, A, Häkkinen, L. 1997. IMAGE magnetometer network, in Satellite-ground based coordination sourcebook, ed. Lockwood M., Wild M.N., Opgenoorth H.J. ESA publications SP-1198, P. 111–117.

Viljanen, A, Pulkkinen, A, Pirjola, R, Pajunpää, K, Posio, P, Koistinen, A. 2006. Recordings of geomagnetically induced currents and a nowcasting service of the Finnish natural gas pipeline system. *Space Weather*, **4**: S10004. https://doi.org/10.1029/2006SW000234.

Viljanen, A. 2011. European Project to Improve Models of Geomagnetically Induced Currents. *Space Weather*, **9**: S07007. https://doi.org/10.1029/2011SW000680.

Viljanen, A, Pirjola, R, Wik, M, Ardarm, A, Prarcser, E, Sakharov, Y, Katkalov, J. 2012. Continental scale modelling of geomagnetically induced currents. *J. Space Weather Space Clim.*, **2**: A17. https://doi.org/10.1051/swsc/2012017.

Vorobjev, VG, Sakharov, YaA, Yagodkina, OI, Petrukovich, AA, Selivanov, VN. 2018. Geomagnetically induced currents and their relationship with locations of westward electrojet and auroral precipitation boundaries. *Tr. Kol'sk. Nauchn. Tsentra Ross. Akad. Nauk*, **4**: 16–28. DOI: 10.25702/KSC.2307-5252.2018.9.5.16-28 (in Russian).

Watari, S, Nakamura, S, Ebihara, Y. 2021. Measurement of geomagnetically induced current (GIC) around Tokyo, Japan. *Earth, Planets and Space*, **73**: 102. https://doi.org/10.1186/s40623-021-01422-3.

Yagova, NV, Pilipenko, VA, Sakharov, YaA, Selivanov, VN. 2021. Spatial scale of geomagnetic Pc5/Pi3 pulsations as a factor of their efficiency in generation of geomagnetically induced currents. *Earth, Planets and Space*, **73**: 88. https://doi.org/10.1186/s40623-021-01407-2.

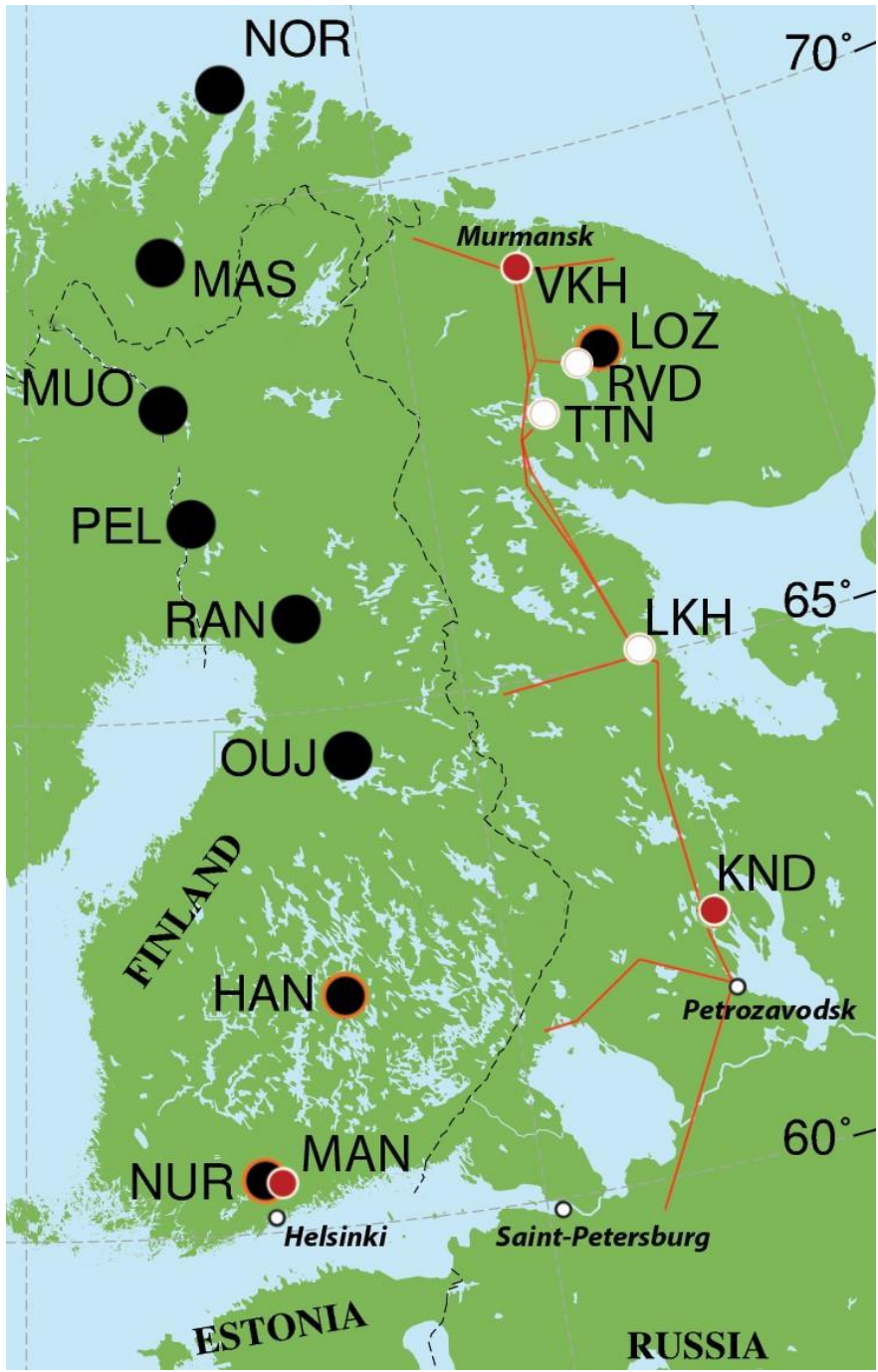

**Figure 1.**

The geographic map of GICs recording stations and nearby observatories for measuring geomagnetic variations both in the north-west of Russia and in Finland. Black dots – IMAGE magnetometer stations, red dots with a white border – used in work GIC registration stations, black dots with an orange border – the magnetometer stations closest to GIC registration stations, white dots – GIC registration stations with no data at selected time.

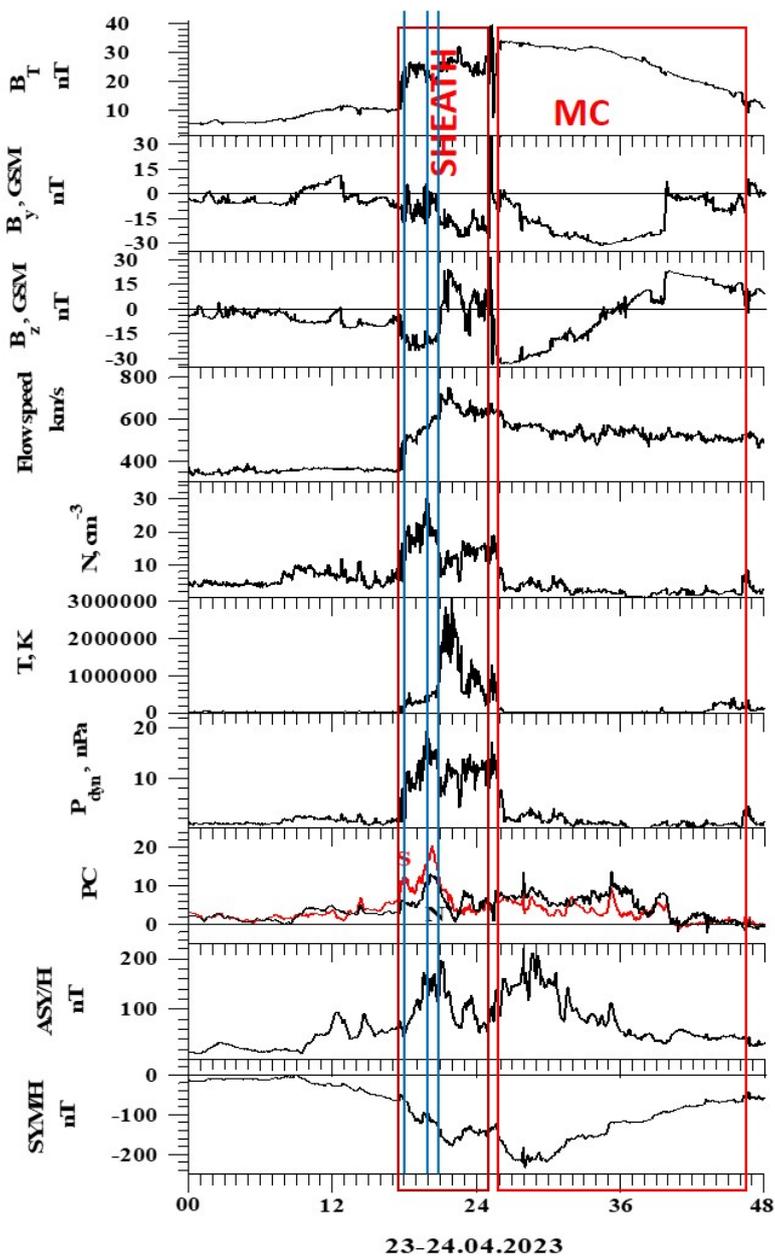

**Figure 2.**

Solar wind and interplanetary magnetic field (IMF) parameters from 00 UT on 23 April to 24 UT April 24, 2023. From top to bottom: the magnitude ($B_T$) of the interplanetary magnetic field (IMF); the Y- and Z-component of the IMF ($B_Y$, $B_Z$); the flow speed (V), the density (N), temperature (T), and dynamic pressure (P) of solar wind, and the geomagnetic indices *PC*, *ASYM/H* and *SYM/H*. The boundaries of the sheath and magnetic cloud are marked by the red rectangles and inscriptions "Sheath" and "MC", respectively; the onsets of substorms at the IMAGE chain are market by blue vertical lines.

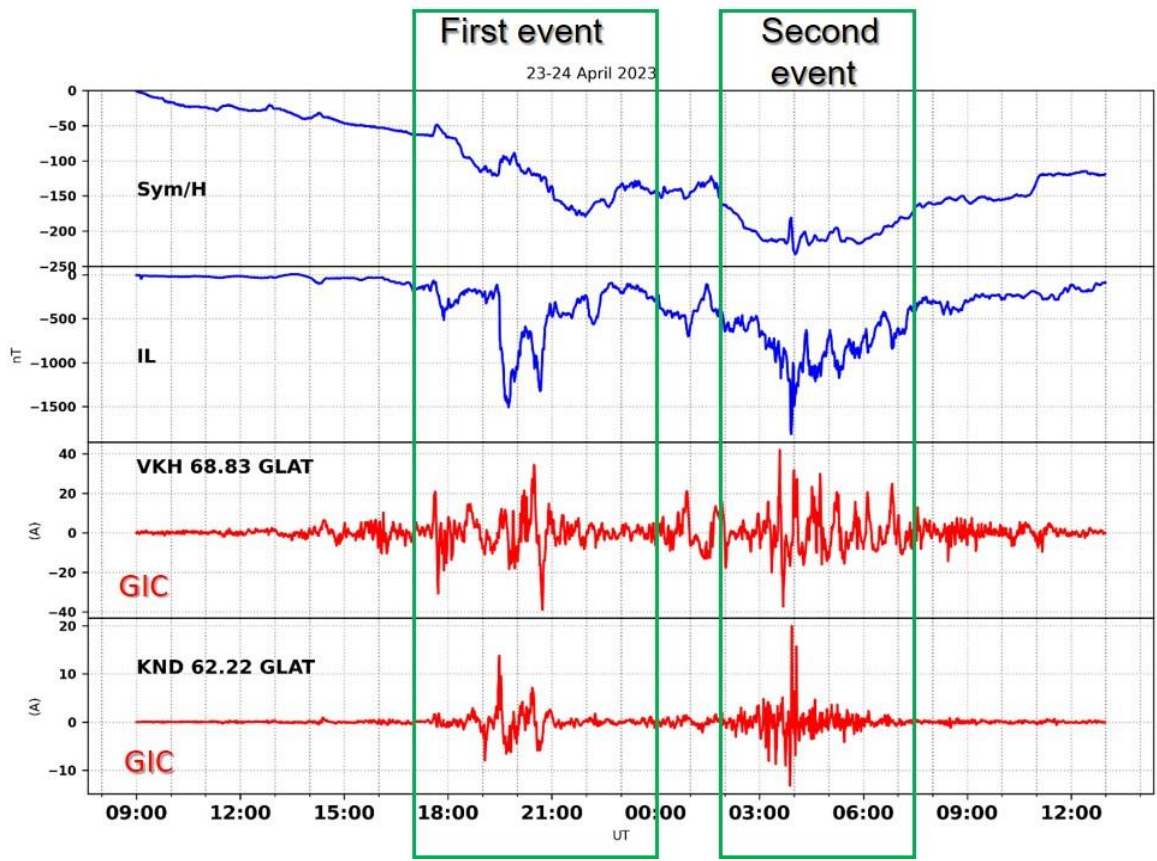

**Figure 3.**

Variations of the geomagnetic indexes and GICs registrations from 09 UT on 23 April to 12 UT on 24 April, 2023. Two top panels show SYM/H and IL indexes (blue lines), two bottom panels - GIC data at Vykhodnoy (VKH) and Kondopoga (KND) (red lines). The first and second magnetic storms are marked by green rectangles and inscriptions "First interval" and "Second interval" respectively.

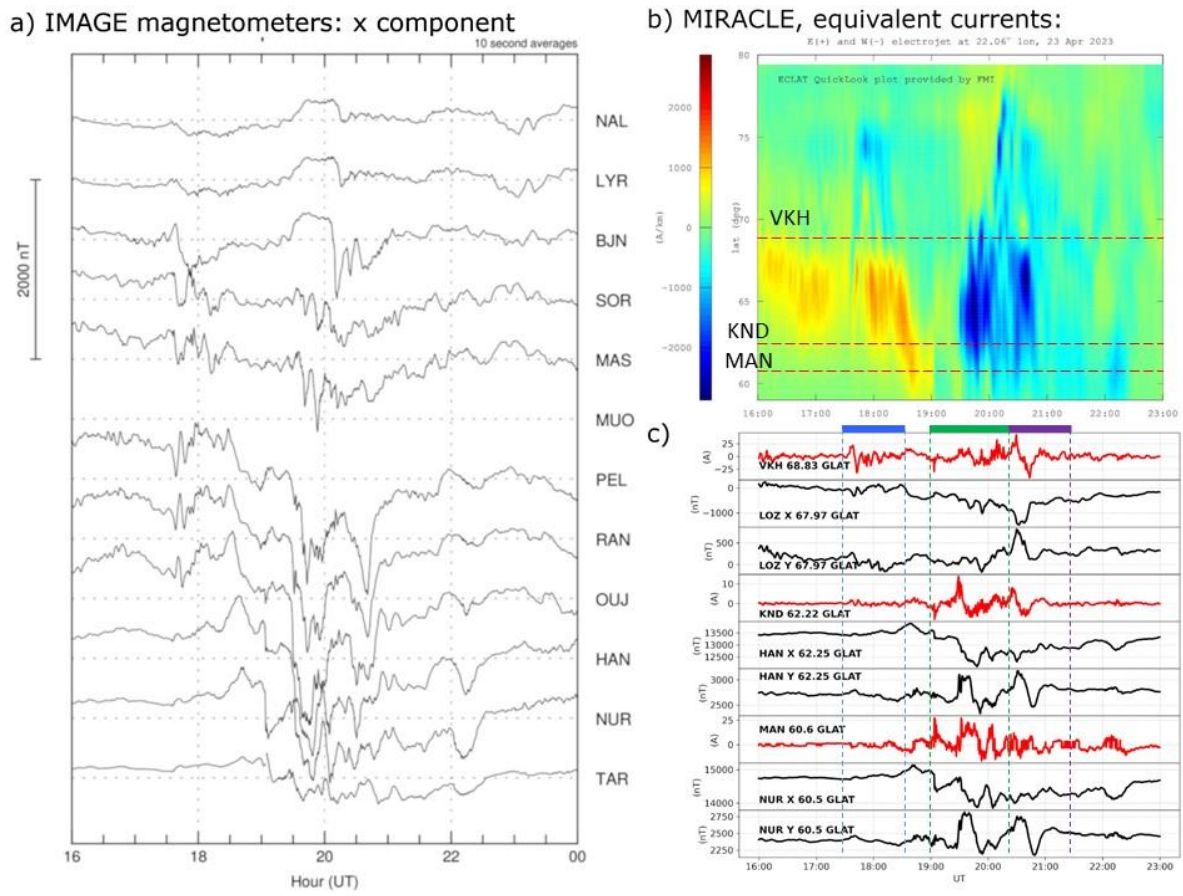

**Figure 4.**

The first interval from 16 UT to 24 UT on 23 April 2023; a) the X component of TAR-NAL chain of IMAGE magnetometers from 16 UT to 24 UT. b) latitudinal profile of the westward (blue) and eastward (yellow) electrojets calculated by MIRACLE system from 16:00 UT to 23:00 UT, the location of VKH, KND and MAN stations are marked by horizontal red dashed lines; c) GICs data (red lines) with corresponding X- and Y-component of nearest IMAGE station (black lines) from 16:00 UT to 23:00 UT. The period of the first, second and third substorms are marked by blue, green and purple bars, respectively.

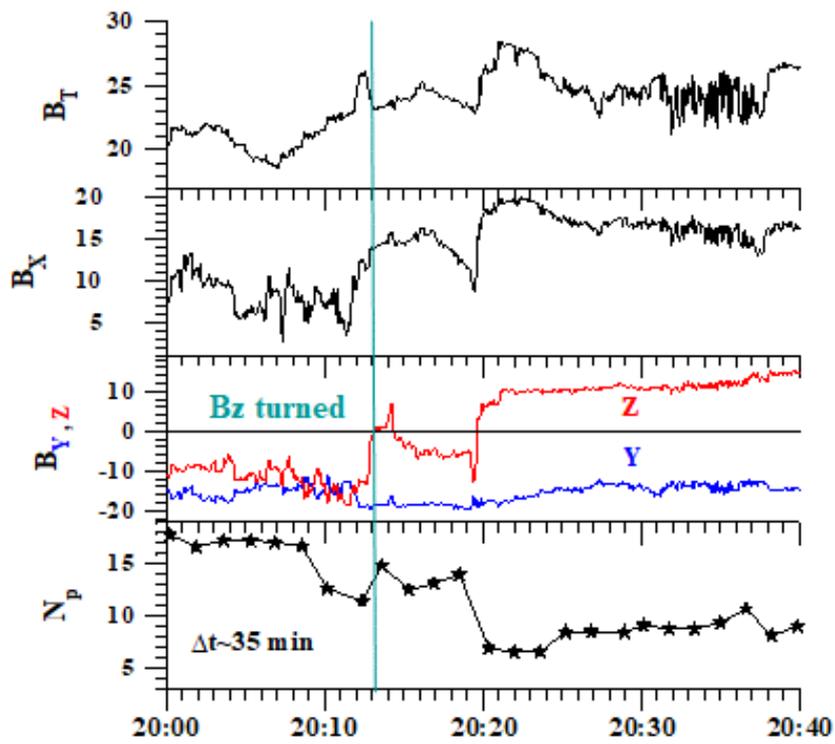

**Figure 5.**

Solar wind and IMF conditions from 20:00 to 20:40 UT on April 23, 2023. From top to bottom: the magnitude $B_T$, $B_X$, $B_Y$ and $B_Z$ components of the IMF and proton density ($N_P$). The time delay from WIND satellite to Earth's orbit is marked by inscriptions on the bottom panel.

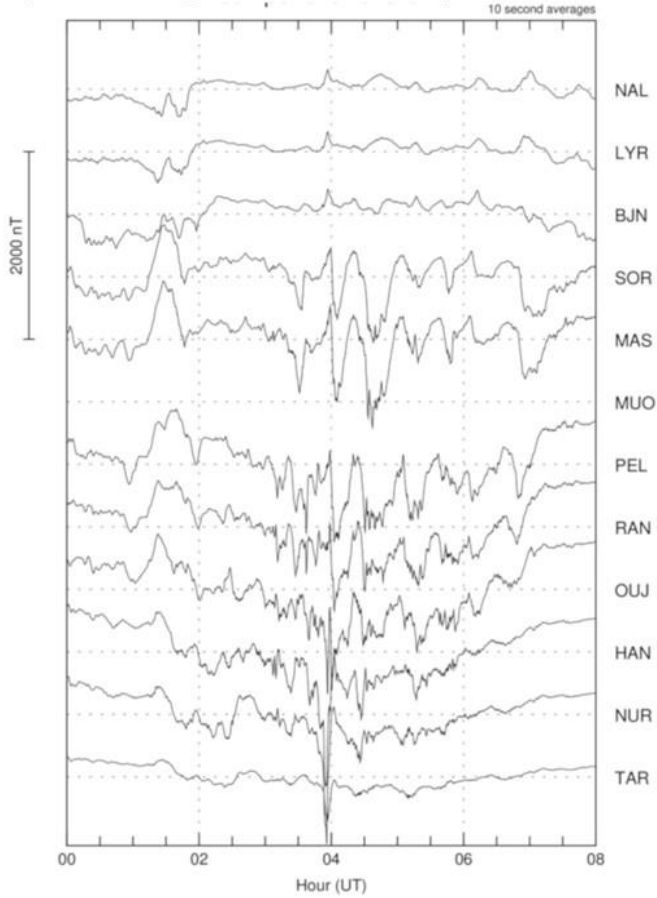
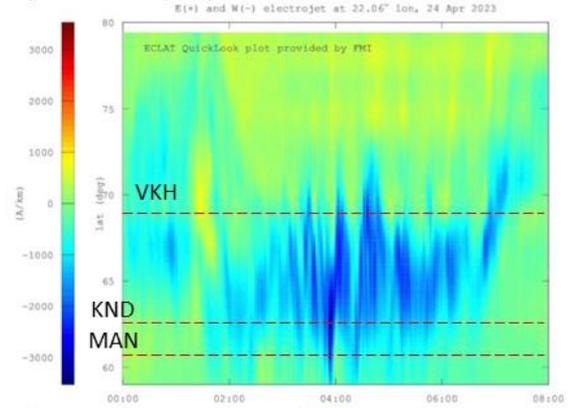
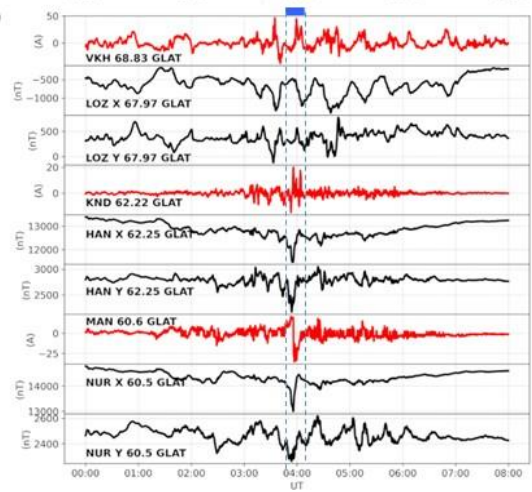

**Figure 6.**

The second interval from 00 to 08 UT on April 24, 2023. The format of Figure 6 is similar to Figure 4. The period of the substorm-like disturbance is marked by blue bar.

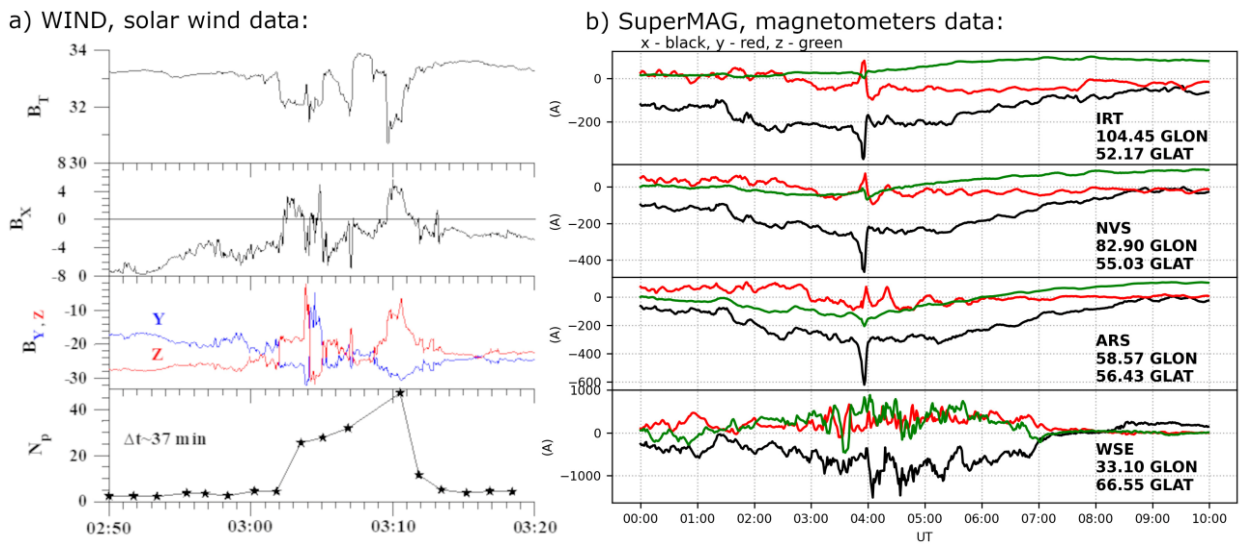

**Figure 7.**

a) Solar wind and IMF conditions from 02:50 UT to 03:20 UT on April 24, 2023. The format of Figure 7a is similar to Figure 5.

b) SuperMAG data from 00 UT to 10 UT on April 24, 2023. Variations of X- (black line), Y- (red line), Z- (green line) components of the magnetic field at two Siberian stations (IRT, NVS), at one Ural station (ARS) and one station located at the north of Karelia (WSE).

**Table 1.**

A list of the coordinates and names used in work IMAGE magnetometers and GICs stations (in bold). The magnetometer stations nearby to GICs stations are underline and in italic.

| Station Code | Station Name | GGLAT (deg) | GGLON (deg) | GGMLAT (deg) | GGMLON (deg) |
|---|---|---|---|---|---|
| NAL | Ny Ålesund | 78.92 | 11.95 | 75.25 | 112.08 |
| LYR | Longyearbyen | 78.20 | 15.82 | 75.12 | 113.00 |
| HOR | Hornsund | 77.00 | 15.60 | 74.13 | 109.59 |
| BJN | Bear Island | 74.50 | 19.20 | 71.45 | 108.07 |
| SOR | Sørøya | 70.54 | 22.22 | 67.34 | 106.17 |
| MAS | Masi | 69.46 | 23.70 | 66.18 | 106.42 |
| **VKH** | **Vykhodnoy** | **68.83** | **33.08** | **65.53** | **112.73** |
| MUO | Muonio | 68.02 | 23.53 | 64.72 | 105.22 |
| *LOZ* | *Lovozero* | *67.97* | *35.02* | *64.19* | *114.46* |
| PEL | Pello | 66.90 | 24.08 | 63.55 | 104.92 |
| RAN | Ranua | 65.90 | 26.41 | 62.09 | 105.91 |
| OUJ | Oulujärvi | 64.52 | 27.23 | 60.99 | 106.14 |
| *HAN* | *Hankasalmi* | *62.25* | *26.60* | *58.69* | *104.54* |
| **KND** | **Kondopoga** | **62.22** | **33.08** | **58.89** | **109.51** |
| **MAN** | **Mäntsälä** | **60.60** | **25.20** | **57.29** | **102.05** |
| *NUR* | *Nurmijärvi* | *60.50* | *24.65* | *56.89* | *102.18* |
| TAR | Tartu | 58.26 | 26.46 | 54.47 | 102.89 |
| BRZ | Birzai | 56.17 | 24.86 | 52.30 | 100.81 |
| SUW | Suwałki | 54.01 | 23.18 | 49.97 | 98.70 |